%% file: soliton_gate.tex
\documentclass[aps,pra,twocolumn,groupedaddress,showpacs]{revtex4-1}
\usepackage{amsfonts}
\usepackage{amsmath}
\usepackage{amssymb}
\usepackage{graphicx}
\usepackage{epsf}
\usepackage{epsfig}
\usepackage{hyperref}
\usepackage{tabularx}

\newcommand{\be}{\begin{equation}}
\newcommand{\ee}{\end{equation}}
\newcommand{\bem}{\begin{multline}}

\begin{document}



\title{Entanglement Generation Using Discrete Solitons in Coulomb Crystals}

\author{H. Landa$^{1,2}$, A. Retzker$^3$, T. Schaetz$^{4}$ and B. Reznik$^1$}
\affiliation{$^1$School of Physics and Astronomy, Raymond and Beverly Sackler Faculty of Exact Sciences, Tel-Aviv University, Tel-Aviv 69978, Israel \\
$^2$Univ.~Paris Sud, CNRS, LPTMS, UMR 8626, Orsay 91405, France\\
$^3$Racah Institute of Physics, The Hebrew University of Jerusalem, Jerusalem 91904, Givat Ram, Israel\\
$^4$Albert-Ludwigs-Universitaet Freiburg, Physikalisches Institut, Hermann-Herder-Strasse 3, 79104 Freiburg, Germany}


\begin{abstract}

Laser cooled and trapped ions can crystallize and feature discrete solitons, that are nonlinear, topologically-protected configurations of the Coulomb crystal. Such solitons, as their continuum counterparts, can move within the crystal, while their discreteness leads to the existence of a gap-separated, spatially-localized motional mode of oscillation above the spectrum. Suggesting that these unique properties of discrete solitons can be used for generating entanglement between different sites of the crystal, we study a detailed proposal in the context of state-of-the-art experimental techniques.
We analyze the interaction of periodically-driven planar ion crystals with optical forces, revealing the effects of micromotion in radiofrequency traps inherent to such structures, as opposed to linear ion chains. The proposed method requires Doppler cooling of the crystal and sideband cooling of the soliton's localized modes alone. Since the gap separation of the latter is nearly independent of the crystal size, this approach could be particularly useful for producing entanglement and studying system-environment interactions in large, two- and possibly three-dimensional systems.

\end{abstract}


\maketitle

\begin{figure}[ht]
\center {\includegraphics[width=3.0in]{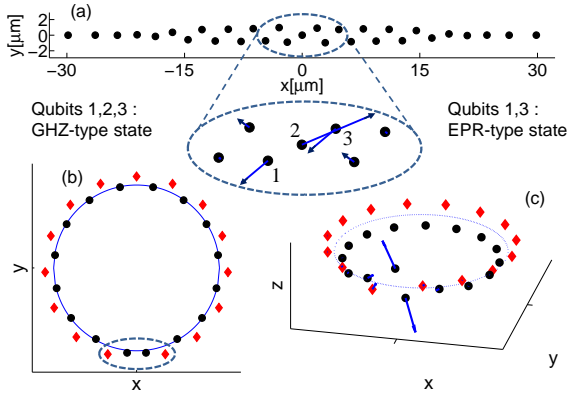}
\caption
{Discrete soliton configurations (numerically simulated) in different radiofrequency traps with 31 ions, the arrows depicting the components of the high-frequency localized normal mode. (a) A planar crystal in a linear trap (top-view), (b) a two-species planar crystal (top-view), which can form in a quadrupole trap with a ring geometry (using e.g. $^{40}$Ca$^+$ and $^{43}$Ca$^+$), or in a linear multipole trap with ions of significantly different masses, e.g. Ca$^+$ and Mg$^+$ [see also Fig.~\ref{Fig:RingScheme}(d)], (c) two rings in a multipole trap (bird's eye view). The crystals in (b) and (c), with a total odd number of ions, are the global minimum configuration and are topologically protected. The lighter ions (black circles) reside at a smaller radius and used as qubit. Heavier ions (red diamonds) are used for sympathetic cooling and structural manipulations.
\label{Fig:Kinks}}}
\end{figure}

The concept of entanglement is crucial for the study of many-body quantum phenomena, and is a key ingredient in quantum computation, simulation, communication and metrology \cite{ladd2010quantum,cirac2012goals,bouwmeester2010physics,PhysRevLett.96.010401,Chou24092010}. Significant progress has been achieved in experiments with a wide variety of solid-state and atomic systems \cite{Devoret08032013,Awschalom08032013,ritter2012elementary,bloch2012quantum}, and extraordinary control has been demonstrated with systems of trapped ions \cite{Leibfried04062004,benhelm2008towards}. At sufficiently low temperatures the ions self-arrange in a Coulomb crystal which can take different forms, e.g. a one-dimensional (1D) chain or a quasi-2D zigzag configuration \cite{Peeters1D,MorigiFishman2008structural}. Proof-of-principle experiments of basic quantum information processing in 1D ion crystals, with up to $\sim 15$ ions \cite{PhysRevLett.106.130506}, have set the state-of-the-art in the field, and scaling up to larger numbers and 2D systems will be the next important challenge \cite{Monroe08032013,1367-2630-15-8-085009}.

As suggested in \cite{KinkCoherence}, stable topological defects in zigzag-shaped ion crystals manifest properties of solitons (kinks), extensively studied in, e.g., sine-Gordon field theory \cite{Rajaraman} and the discrete Frenkel-Kontorova model \cite{BraunKivsharBook}. Solitons carry localized collective motional ``internal modes'' bound to the soliton, which have been proposed for carrying quantum information and entanglement \cite{Marcovitch}, while entanglement in macroscopic coordinates of colliding solitons has been studied in \cite{LewensteinMalomed}. Discrete solitons in zigzag crystals can be spatially extended configurations or highly localized with a size of a few ions. In the latter case they manifest a high-frequency localized motional mode, spectrally separated from the other phonons by a gap that is nearly independent of the crystal size \cite{KinkCoherence} -- a property which plays a key role in the present work. 
Recently such structures have been experimentally created, positioned and stabilized in linear Paul traps \cite{Schneider2012,kink_trapping,TanjaKZ,MainzKZ,HaljanKinks,KinkBifurcations,TanjaPN}. In linear multipole traps \cite{GerlichChem,OctupoleCrystalExp,OctupoleCrystalExp2} and in circular traps \cite{ToroidalTrap,SchaetzCrystallineBeams,PhysRevE.66.036501}, ions can crystallize in one or more rings, which can be advantageous for studying various effects \cite{maleki2004search,MartinaRing,AcousticBlackHole,PhysRevLett.109.163001,nigmatullin2011formation}, in particular discrete solitons \cite{KinkCoherence,multipole_kinks,MorigiMultipoleRings}. 

In this Letter, we propose to employ discrete solitons for the generation and distribution of entanglement within a system of many particles in a thermal motional state. The key ingredients of our proposal are as follows: (i) Two- and multi-qubit entangled states are generated locally at the position of the soliton, utilizing the gapped soliton modes. (ii) To entangle other qubits, the soliton can be translated deterministically along the crystal with step (i) repeated, thereby allowing remote qubits to be sequentially entangled. (iii) Circular crystal topologies provide a flexible setup for combining cooling, entangling and moving the soliton in a robust manner.

We study in detail the above ideas in the setup of a crystal of trapped ions. Steps (i)-(ii) can be implemented in linear Paul traps with existing technology. Entangling a few ions within a large crystal necessarily requires to decouple them from the rest of the crystal \cite{sackett2000experimental,Rydberg_mode_shaping}. Based on the localized nature of the kink modes, collective laser addressing suffices for entangling electronic qubit levels with a small subset of ions, without the need to spatially move them or cool the entire crystal close to the motional ground state, as the gap-separation of the high-frequency mode allows to spectrally resolve it within a dense spectrum of modes. Circular crystal topologies [(iii)] can be realized in ring and linear multipole ion traps, and in particular a two-species crystal [fig.~\ref{Fig:Kinks}(b)-(c)] can be advantageous for control of the soliton motion and continuous cooling during the gate operation.

\textit{Entanglement generation.} We first consider a single kink within a planar zigzag crystal of a few tens of ions in a linear Paul trap. Fig.~\ref{Fig:Kinks}(a) shows a zoom-in on the center of the kink configuration, with the arrows depicting the ion components of the high-frequency vibrational normal mode to be employed as the localized `quantum bus' used to entangle qubits in the S{\o}rensen M{\o}lmer (SM) gate \cite{sorensen1999quantum,Roos2008}. In the following we will show that a wide laser beam is sufficient for addressing only the qubits stored in the three center ions (denoted 1,2,3) without coupling to the other ions. The unitary transformation implemented on these qubits is, up to local qubit phases,
\be U_{\rm eff}=\exp\{i\frac{\pi}{8\alpha} S^2\},\qquad S\equiv \tilde{\sigma}_{y,1}+ \alpha\tilde{\sigma}_{y,2}+ \tilde{\sigma}_{y,3},\label{eq:U_eff}\ee
where $\tilde{\sigma}_{y,i}\equiv\sigma_{y,i}\cos\tilde{\phi}_i-\sigma_{x,i}\sin\tilde{\phi}_i$ are Pauli matrices acting on qubit $i$, with $\tilde{\phi}_i$ depending on the position of each ion and $\alpha \neq 0$ depends on the contribution of the bus mode, and is also affected by the amplitude of the radiofrequency (rf) micromotion (see eq.~\eqref{Eq:alpha}). Starting with the initial state $\left|ggg\right>$ (where $\left|g\right>$ and $\left|e\right>$ denote the electronic qubit states), $U_{\rm eff}$ creates a GHZ-type state of 3 qubits (whose 3-tangle measure \cite{PhysRevA.61.052306,PhysRevA.62.062314} equals 1), $\left|ggg\right>+e^{i\theta_1}\left|gee\right>+e^{i\theta_2}\left|ege\right>+e^{i\theta_3}\left|eeg\right>$, where $\theta_i$ are determined by $\phi_i$ defined below, and $\left[U_{\rm eff}\right]^2$ creates an entangled state of qubits 1 and 3 (maximally entangled for $\alpha=2$), with qubit 2 disentangled.

\begin{figure}[ht]
\center {\includegraphics[width=3.0in]{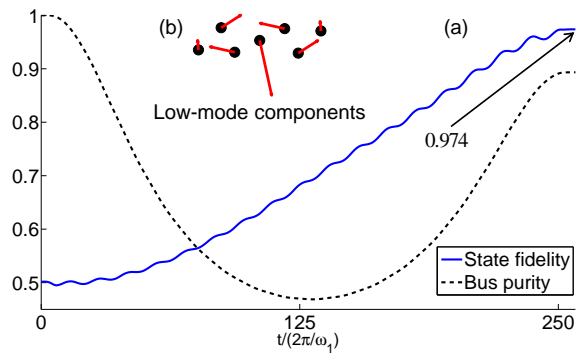}
\caption
{(a) Fidelity (${\rm tr}\sqrt{\sqrt{\rho}\rho_{\rm ideal}\sqrt{\rho}}$) due to nonlinear mode heating during the generation of the entangled GHZ-type state of 3 qubits within a 31-ion crystal in a linear Paul trap, with respect to the ideal final state, from an integration of the master equation in eq.~\eqref{eq:Master} using the exact Hamiltonian of eq.~\eqref{eq:H_i}, when cooling the high- and the low-frequency localized modes to the ground state. A measure for the purity (${\rm tr}\rho_{\rm bus}^2$) of the bus mode is also presented -- in the ideal case it would return to its initial pure state at the end of the gate (after $\sim 250$ mode oscillations). (b) The spatial components of the in-plane low-frequency localized mode.}
\label{Fig:Fidelity}}
\end{figure}

\textit{Micromotion treatment.} Taking into account the time-dependent potential of Paul traps, the coordinates of ion $i$ can be expanded as $\vec{R}_{i}\left(t\right)=\vec{R}_{i}^{\pi}\left(t\right)+\delta \vec{R}_{i}\left(t\right)$. The first term is the classical driven solution which is the dynamical equivalent of a minimum-configuration location \cite{rfions}. The ions oscillate about an average position at the rf frequency, which after conventional rescaling of time, becomes $\Omega_{\rm rf}\equiv 2$, such that $\vec{R}_i^\pi\left(t\right)=\sum_{n=-\infty}^{\infty}\vec{B}_{2n,i}e^{2int}$ is $\pi$-periodic with $\vec{B}_{2n,i}$ constants. The linearized ion deviations from the periodic solution, $\delta \vec{R}_{i}\left(t\right)$, are canonically quantized and can be expressed in terms of linear combinations of creation and annihilation operators of decoupled phonons by using a (`Floquet-Lyapunov') time-dependent transformation \cite{rfions,rfmodes}.

The interaction Hamiltonian of ion qubit $i$, coupled by the laser to the phonon mode $j$, is given in the interaction picture and rotating wave approximation (with respect to the laser), by
\be H_i = \frac{1}{2}\hbar \Omega_i\left(t\right)e^{-i\phi_i\left(t\right)}\sigma_{+,i}e^{i\eta_j\left\{\lambda_i^j\left(t\right)b_j^\dagger\left(t\right)+\lambda_i^j\left(t\right)^*b_j\left(t\right) \right\}}+\rm{h.c.},\label{eq:H_i}\ee
where $\sigma_{+,i}$ is the qubit raising operator, $b_j\left(t\right)=b_j\left(0\right) e^{-i\omega_j t}$ is the phonon annihilation operator and $\eta_j$ is the Lamb-Dicke parameter of the phonon whose secular frequency is $\omega_j$. The time-dependence of the trapping potential enters into eq.~\eqref{eq:H_i} in 3 ways (as compared with a pseudopotential approach), which we analyze concisely \cite{supplemental}. The secular frequencies $\omega_j$ are shifted, which we account for by using the exact mode frequencies in rf traps \cite{rfions,zigzagexperiment}. The periodic modulation of the laser's wavefront phase at the ion's position, given by $\phi_i\left(t\right)=\phi_L-\vec{k}\cdot\vec{R}_i^\pi\left(t\right)$, with $\phi_L$ the laser's optical phase and $\vec{k}$ its wavevector, leads to a significant effect on gates in the current setting, as detailed below. The mode coefficients $\lambda_i^j\left(t\right)$ which appear in eq.~\eqref{eq:H_i} are in general periodic functions of time, but can be replaced by the leading order (constant) term $\tilde{\lambda}_i^j$ in the case of the bus mode \cite{supplemental}.

For the SM gate using bichromatic copropagating laser beams we set for the Rabi frequency of the interaction, $\Omega_i\left(t\right)=\Omega \left(e^{i\delta t}+e^{-i\delta t}\right)f\left(t\right)$ where $\Omega$ can be tuned by the laser intensity (the laser profile assumed to be uniform), $f\left(t\right)$ is a window function smoothly turning on and off the interaction \cite{Roos2008}, which must be synchronized with the phase of the rf trapping potential, and $\delta$ is the laser detuning from the qubit transition frequency. With the Lamb-Dicke coefficient of order $10^{-2}$, we can expand the Hamiltonian of eq.~\eqref{eq:H_i} and obtain to first order
\begin{multline} H_I \approx \sum_i\frac{1}{2}\hbar \Omega_i\left(t\right){J}_0\left(2\vec{k}_y \vec{B}_{2,i,y}\right) \times \\\times \left[ \tilde{\sigma}_{x,i} -\eta_j\tilde{\lambda}_i^j\left(b_j^\dagger\left(t\right)+ b_j\left(t\right) \right)\tilde{\sigma}_{y,i}\right],\label{eq:H_i2}\end{multline}
where each ion's phase within the laser has been absorbed into the Pauli matrices rotated in $xy$ qubit-plane by the angle
$\tilde{\phi}_i\equiv\phi_L-\vec{k}\cdot \vec{B}_{0,i}$, as defined following eq.~\eqref{eq:U_eff}. The Bessel function factor accounts for the phase-modulation caused by each ion's micromotion oscillation \cite{PhysRevLett.81.3631,supplemental}.

To avoid any significant processes involving other modes, the difference of the laser detuning from the bus mode frequency ($\epsilon=\omega_1-\delta$, where $j=1$ is the high-frequency mode) is chosen to be much smaller than the gap to the other phonons (the gap is $\sim 1/12$ of $\omega_1$, see below). This requires to terminate the gate operation after (multiples of) $\omega_1/\left|\epsilon\right|$ mode periods to get the qubits in a pure state disentangled from the phonon.
The effective interaction induced by $H_I$ at time of first disentanglement of the phonon from the qubits, $t^*=2\pi/\left|\epsilon\right|$, is given by $U_I\left(t^*\right)=\exp\{i\theta t^* S^2\}$ with $S$ defined in eq.~\eqref{eq:U_eff}, and choosing the parameters such that $\theta t^*=\pi/8\alpha$ gives $U_{\rm eff}$. The amplitude factor of ion 2 in eq.~\eqref{eq:U_eff} is then
\be\alpha=\tilde{\lambda}_2^j/\left[\tilde{\lambda}_1^j{J}_0\left(2\vec{k}_y \vec{B}_{2,1,y}\right)\right],\label{Eq:alpha}\ee
where, since the middle ion performs no micromotion, $J_0\left(0\right)=1$ in the nominator of eq.~\eqref{Eq:alpha}.
If the micromotion phase-modulation \cite{supplemental} did not exist, the phonon components could be chosen such that $|\alpha|=2$, with $\left[U_{\rm eff}\right]^2$ creating a maximally-entangled EPR-pair. However, the rf modulation turns out to be nonnegligible and for the chosen parameters gives $|\alpha|=2.22$, so the resulting pure state is approximately $\frac{1}{\sqrt{2}}\left(0.92\left|gg\right>+1.075i\left|ee\right>\right)$.

\textit{Simulation parameters}. We consider $^{40}{\rm Ca}^+$ ions with the dipole-forbidden transition at $729$~nm serving as qubit. The secular trapping frequencies (with the axes defined in fig.\ref{Fig:Kinks}) are $\omega_x=2\pi\cdot 700~{\rm kHz}$, $\omega_y=8.38\omega_x\approx 2\pi\cdot 5.87~{\rm MHz}$ and $\omega_z/\omega_y=1.16$. The rf frequency is $\Omega_{\rm rf}=2\pi\cdot 80.8~{\rm MHz}$ and the Mathieu parameter $q_y= 0.22$. The high-frequency mode has $\omega_1=12.02\omega_x\approx 2\pi\cdot 8.42~{\rm MHz}$, the gap is $\omega_1/\omega_2\approx 1.085$ or equivalently $\omega_1-\omega_2=0.95\omega_x\approx 2\pi\cdot 0.66~{\rm MHz}$, and $\epsilon \equiv \omega_1-\delta = 0.004\omega_1$ ($\delta$ being the laser detuning from the electronic transition frequency), making $\epsilon \sim 1/20$ of the gap. The low frequency mode has $\omega_{\rm low}=1.47\omega_x\approx 2\pi\cdot 1.03~{\rm MHz}$. The Rabi frequency is $\Omega= 0.1225\omega_1 \approx 2\pi\cdot 1.03~{\rm MHz}$. The Lamb-Dicke coefficient multiplied by the laser projection on the bus mode component for ion 1 gives $\eta_1\tilde{\lambda}_1^1\approx 0.0121$, and the factor on the middle qubit is according to eq.~(4), $\alpha=-0.0237/\left[-0.0121\cdot J_0\left(-2\cdot 0.35\right)\right]\approx -2.22$. The laser propagation direction is chosen such that $\tilde{\lambda}_i^1 \lesssim 1/100$ for the bus mode on all ions except the 3 relevant ones, and $\tilde{\lambda}_i^j\lesssim\tilde{\lambda}_i^1/5$ for those ions on off-resonant modes ($j=2,3$, the only ones capable to contribute), making the effective relative gate strength on those modes of order $10^{-3}$.

\begin{table}
	\centering
		\begin{tabularx}{\columnwidth}{c X X X X X}
		\hline\hline
		Number of ions & 31 & 61 & 91 & 121 & 151 \\
		\hline
		~~ Rate $\left(\cdot 10^{-4}\right)~~ $ & 0.75 & 2.2 & 2.0 & 2.0 & 1.8\\ \hline\hline
		\end{tabularx}
	\caption{Rate of nonlinear heating of the bus mode from the ground state, as function of the number of ions in the crystal, showing the increase of the average phonon number, per oscillation period of the mode, due to its coupling to the bath of all other modes, calculated using a non-Markovian master equation (see text). The radial trapping frequency was held fixed while the axial frequency was decreased with ion number to give configurations with similar spatial and spectral kink properties. The parameters were only optimized for the case of 31 ions.}	\label{tab:HeatingRates}
\end{table}

\textit{Open quantum system effects.} The bus mode is coupled to the other crystal modes (viewed as a thermal bath) mostly through 3-phonon processes which involve the localized low-frequency mode [fig.~\ref{Fig:Fidelity}(b)] and another mode from the bath \cite{KinkCoherence}. We first integrate the non-Markovian master equation \cite{KinkCoherence} for the bus mode coupled to the bath of all other modes, taking different initial states. We observe recoherences of the bus mode state which can be controlled by tuning the nonlinear resonances of the localized modes. Such aspects are promising for investigating open quantum system dynamics \cite{PhysRevA.87.050304,nonMarkovian2013}. By cooling the localized modes close to the ground state, the strength of the nonlinear processes can be significantly decreased. This is technically feasible since the localized modes are gapped and hence individually addressable (the kink has one in-plane and one out-of-plane low-frequency modes, whose frequencies can be controlled by varying $\omega_y$ and $\omega_z$ respectively). Thereby the nonlinear coupling of the bus mode can be approximated by a small effective heating rate of the mode from the ground state, which we extract (conservatively) and then in the second step plug into the Markovian master equation
\be \dot{\rho}\left(t\right)=-\frac{i}{\hbar}\left[\sum_i H_i\left(t\right),\rho\left(t\right)\right]-\mathcal{L}\left[\rho\left(t\right)\right], \label{eq:Master} \ee
where $\rho$ is the combined density matrix of the qubits and the phonon, $H_i$ is defined in eq.~\eqref{eq:H_i} and $\mathcal{L}\left[\rho\left(t\right)\right]$ is a Lindbladian part \cite{carmichael1993open} which introduces the effective heating of the bus mode. The fidelity of the entangled state creation (calculated using \cite{machnes2007qlib}) is depicted in fig.~\ref{Fig:Fidelity}.

\begin{figure}[ht]
\center {\includegraphics[width=3.2in]{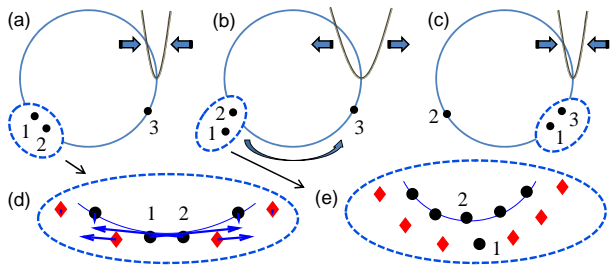}
\caption
{Entangling remote pairs of ions in circular crystals. Starting with a localized soliton configuration [(a)], similar to that of fig.~\ref{Fig:Kinks}(b), qubits ions 1 and 2 are entangled using the localized bus mode [(d)]. Decreasing the radial trapping frequency, an extended soliton is formed [(b)], in which one of the two qubit ions (black circles) is pushed into the outer ring of the wide zigzag. The kink in the outer ring can be pushed to `slide' over the internal ring, carrying with it qubit ion 1, one lattice site [(e)]. The radial trapping can be increased and the kink be pinned after having reached a desired position [(c)], with its modes cooled and used to perform a gate between qubit 1 and its new neighbor, qubit 3.}
\label{Fig:RingScheme}}
\end{figure}

Examining the evolution without the bath, the 3 qubits and the bus mode become entangled, and the average phonon number grows up to 0.78 at $t^*/2$. At $t^*$, the qubits and phonon disentangle and the phonon returns (ideally) to the ground state. For the chosen parameters the 3-qubit entangled state can be generated with about 97.4\% fidelity (fig.~\ref{Fig:Fidelity}), and an entangled pair can be created in qubits 1 and 3 at time $2t^*$ with nearly twice the infidelity. The heating rate depends only weakly on the number of ions (see Table \ref{tab:HeatingRates}), and the resulting entangling infidelity remains roughly constant for 60-150 ions; with 151 ions a fidelity of 93.3\% was simulated (without a full optimization of the parameters). On the other hand, the nonlinear coupling drops with increasing the trapping frequencies, due to lower excitations of the bath modes; when doubling the axial trap frequency to $\omega_x=2\pi\cdot 1400{\rm kHz}$ (and scaling all other frequencies accordingly), the GHZ-type state can be created with a fidelity of 98\% (with 31 ions).

\textit{Circular topologies.} In circular crystals the kink is topologically protected for an odd number of ions. Fig.~\ref{Fig:RingScheme} depicts a method for entangling remote qubits in a crystal of two different ion species. The light ions of the inner ring (black circles) are used to store logical qubits, while the heavier ions in the outer ring (red diamonds) are used for sympathetic cooling \cite{morigi2001twospecies,PhysRevA.68.042302} and for controlling the motion of the soliton. By varying the radial trapping potential, a highly localized kink can be transformed into an extended one \cite{TanjaKZ,KinkBifurcations,TanjaPN}, with qubit ion 1 from the inner ring embedded in the outer ring [fig.~\ref{Fig:RingScheme}(b),~(e)]. The motion of this ion can be controlled by exciting a coherent state \cite{PhysRevLett.103.090504,1367-2630-14-3-035012} of the low-frequency mode, with the phase accurately timed with the variation of the trapping potential, which also puts the soliton into motion in the crystal, as we demonstrate by numerically simulating the classical equations of motion \cite{supplemental}. Furthermore, the low-frequency mode can be continuously cooled during the gate operation, reducing the nonlinear coupling of the modes, and sophisticated ideas for continuous cooling of the bus mode can also be explored \cite{bermudez2012dissipation}. An EPR-pair of the two logic ions 1 and 2 can be generated on the high-frequency bus mode using a tangential laser [fig.~\ref{Fig:RingScheme}(d)], which will have a negligible affect on neighboring qubits ions, whose mode components are orthogonal. Single-qubit rotations can be performed on the logic ion in the outer ring of the extended kink configuration [fig.~\ref{Fig:RingScheme}(e)], using a micromotion sideband \cite{PhysRevLett.81.3631}. In this way, a circular topology incorporating a movable discrete soliton provides an elegant possibility of entangling remote qubits in a large, thermal environment. A full scale algorithm would require higher gate fidelities than calculated above, and it is beyond the scope of the current work to investigate how these may be achieved, possibly by using faster gates which result in smaller decoherence.

\textit{Summary and outlook.} We proposed and analyzed in detail a method for generating entanglement using solitons in a system of $N \approx 30-150$ ions. The approach is largely insensitive to the number of trapped ions. In particular, the gap-separated high frequency internal mode is very weakly dependent on $N$ and furthermore, the cooling requirements are sideband cooling of only the localized modes, while Doppler cooling the crystal. We believe therefore, that a proof-of-principle experiment is feasible within current linear Paul traps. From the fundamental perspective of the system-environment decoherence problem, this could then provide the intriguing possibility of exploring in a controlled fashion the generation of a pure entangled state and its subsequent decoherence mechanism, with non-Markovian effects such as coherence recurrences. Our proposal for transporting entanglement in a lattice by controlling the motion of a nonlinear, topological excitation, within the circular topology implementation, is currently experimentally more challenging. However we hope that together with other proposals involving circular geometries \cite{MartinaRing,AcousticBlackHole,PhysRevLett.109.163001}, it would provoke further research. Finally, it could be interesting to study the present ideas in the context of other systems as well, such as zigzag crystals of electrons in 1D quantum wires \cite{PhysRevLett.110.246802}, dipolar gases \cite{PhysRevA.78.063622}, or dark solitons realized with cold atoms \cite{1751-8121-43-21-213001}.

\begin{acknowledgments}
BR and AR acknowledges the support of the Israel Science
Foundation and the support of the German-Israeli Foundation. BR and TS acknowledge the support of the European Commission (STREP PICC). TS was funded by the Deutsche Forschungsgemeinschaft (SCHA973). AR was funded by carrier integration grant (CIG) no.~321798 IonQuanSense FP7-PEOPLE-2012-CIG. HL acknowledges support by the French government via the 2013-2014 Chateaubriand fellowship of the French embassy in Israel. We thank M. B. Plenio for fruitful discussions, held in particular during the ``2013 Ulm Workshop on Theoretical and Experimental Aspects of Nonlinear Physics in Ion Traps''.
\end{acknowledgments}

\appendix
\section*{Supplemental Material}
\input{SupplementalContent}

\bibliographystyle{hunsrt}

\bibliography{soliton_gate}

\end{document}

%% file: SupplementalContent.tex
\subsection*{Micromotion effects}

As discussed in \cite{rfions} and demonstrated experimentally in \cite{zigzagexperiment}, the time-dependence of the trapping potential in Paul traps may significantly modify the structure and frequencies of the crystal phonons as compared with the pseudopotential approximation. In the kink configuration in the linear Paul trap considered in the paper (with parameters as detailed in the following), we find that the frequency (and also the eigenvector components) of the high-frequency mode are almost unaffected, but the low-frequency mode is lowered in frequency by $\sim 2\%$.

In addition to frequency corrections, the time-dependence of the trapping potential enters the quantum interaction of light with ion crystals through the time-dependent terms $e^{-i\phi_i\left(t\right)}$ and $\lambda_i^j\left(t\right)$, defined in eq.~(2) of the main text. 

Treating the first of these terms, for a planar crystal in a linear Paul trap in which micromotion is along the $y$ axis (dependent on the Mathieu parameter $q_y$), we have
\[ \vec{B}_{2,i,y}\approx \frac{q_y}{4}\vec{B}_{0,i,y},\qquad \vec{B}_{4,i,y}\approx \frac{1}{4}\frac{q_y^2}{4^2}\vec{B}_{0,i,y},\]
where the coefficients $\vec{B}_{2n,i}$ are defined in the main text as the constants in the Fourier expansion of the ion $i$'s coordinate.
The amplitude correction resulting from $B_4$ can be calculated to be of order $10^{-4}$ and is neglected. Expanding $e^{-i\phi_i\left(t\right)}$ we get
\[ e^{-i\phi_i\left(t\right)}=e^{-i\phi_L+i\vec{k}\cdot \vec{B}_{0,i}}\sum_{n=-\infty}^{\infty}i^n{J}_n\left(\vec{k}_y \vec{B}_{2,i,y}\right)e^{2int}, \]
where $\phi_L$ is the laser's optical phase and $\vec{k}$ its wavevector, and $J_n$ is the Bessel function of integer order $n$.
For ions extending out of the rf-null line, the argument of the Bessel function may be of order 1, and therefore the correction to the amplitude could be significant. Terms with $n\neq 0$, rotating with the rf frequency, are very far from resonance, resulting in a relative correction which is of order $10^{-4}$ (or smaller), and can be neglected.

The phonon coefficients are defined by
\[\lambda_i^j\left(t\right)=\sum_{\alpha \in \left\{x,y,z\right\}}\hat{k}_{\alpha}U_{i,{\alpha}}^j\left(t\right),\]
with $U$ defining the Floquet-Lyapunov transformation \cite{rfions,rfmodes} which relates the ions' small oscillations to the phonon \cite{rfions}, and $\hat{k}=\vec{k}/\|\vec{k}\|$. Similarly to above, the contribution of terms in the Fourier expansion of $\lambda_i^j\left(t\right)$ rotating at the rf frequency, is of order $10^{-4}$ and can be neglected, leaving only the leading order term, denoted $\tilde{\lambda}_i^j$ below. \linebreak

\subsection*{Motion of a kink in circular crystal}

The motion of a discrete soliton in the crystal can be manipulated by exciting the lowest frequency mode, which describes oscillations of the kink in the effective potential which pins it at a specific position in the lattice \cite{KinkCoherence,KinkBifurcations}. In the circular topologies of figs.~1 and 3 of the manuscript, this can be done by exciting a coherent state of this mode using the same laser used for cooling the low frequency mode, and since this laser can be made resonant only with the electronic transitions of the heavier, cooling ions (in red), this can be done without impeding the qubits stored in the lighter ions.

With this Supplemental Material, we include two movie files depicting a simulation of the classical equations of motion of ion crystals in a circular quadrupole Paul trap as in fig.~1(b). The radial trapping frequency was chosen for convenience to be $2\pi\cdot 1~{\rm MHz}$. The movie in the file titled `Trapping decrease.wmv' shows the motion of the ions, after a classical state with an amplitude equivalent to 1000 phonons was excited in the low mode of a highly localized kink (with the rest of the modes at the temperature corresponding to the Doppler cooling limit). A rapid decrease of the radial trapping frequency, timed with a specific phase of the mode, causes the desired ion from the kink core (ion \#1 of fig.~3) to move into the external ring of heavier (red) ions, and then the crystal is cooled to remove the excessive energy imparted during the process. We note that the energy excited into the low frequency mode exhibits complicated dynamics at different steps of the shown simulation, dispersing into the other modes and returning into the localized mode at specific times.

In the file titled `Kink motion.wmv', a coherent state of the low frequency mode of the extended soliton configuration is excited (with an amplitude corresponding to 5000 phonons), initiating a motion of the soliton in the crystal, which leads, eventually, to displacement by two lattice sites of the qubit ion in the external ring. In this simulation there is no cooling affected on the crystal, however the energy of the localized mode eventually disperses over the entire crystal.